\documentclass{aa}
\usepackage[utf8]{inputenc}
\usepackage{natbib}
\usepackage{amsmath}
\usepackage{graphicx}
\usepackage{color}

\def\gr{$\gamma$-ray}
\def\gr{$\gamma$-ray}
\def\aap{A\&A}

\def\apjl{AP.J.Lett.}

\begin{document}

\title{Galactic diffuse gamma-ray emission at TeV energy }
\author{
A. Neronov \inst{1,2} \and
D. Semikoz \inst{1} }
 \date{}
\institute{APC, University of Paris, CNRS/IN2P3, CEA/IRFU,  10 rue Alice Domon et Leonie Duquet, Paris, France\and
Astronomy Department, University of Geneva, Ch. d'Ecogia 16, 1290, Versoix, Switzerland}

\label{firstpage}

\abstract
{
Measuring the diffuse Galactic \gr\ flux in the TeV range is difficult for ground-based \gr\ telescopes  because of the residual cosmic-ray background, which is higher than the \gr\ flux by several orders of magnitude. Its detection is also challenging for space-based telescopes because of low signal statistics.
 }
{
We characterise the diffuse TeV flux from the Galaxy using decade-long exposures of the Fermi Large Area Telescope. 
} 
{
Considering that the level of diffuse Galactic emission in the TeV band approaches the level of residual cosmic-ray background, we  estimated the level of residual cosmic-ray background in the {SOURCEVETO} event selection and verified that the TeV diffuse Galactic emission flux is well above the residual cosmic-ray background up to high Galactic latitude regions.
}{ 
We study spectral and imaging properties of the diffuse TeV signal from the Galactic plane. We find much stronger emission from  the inner Galactic plane than in previous HESS telescope estimates (lower bound).  We also find a significant difference in the  measurement of the Galactic longitude and latitude profiles of the signal measured by Fermi and HESS. These discrepancies are presumably explained by the fact that regions of background estimate in HESS have non-negligible \gr\  flux.  Comparing Fermi measurements  with those of ARGO-YBJ, we find better agreement, with the notable exception of the Cygnus region, where we find much higher  flux (by a factor 1.5). We also measure the TeV diffuse emission spectrum up to high Galactic latitude and show that the spectra of different regions of the sky have spectral slopes consistent with $\Gamma=2.34\pm 0.04$, which is harder than the slope of the locally observed spectrum of cosmic rays with energies 10-100 TeV, which produce TeV diffuse emission on their way through the interstellar medium.   We discuss the possible origin of the hard slope of the TeV diffuse emission.   
}
{Fermi/LAT provides reliable measurements of the diffuse Galactic emission spectrum in the TeV range, which are almost background free at low Galactic latitudes. The diffuse flux becomes comparable to the residual cosmic-ray background  at Galactic latitudes $|b|>50^\circ$. Its measurement in these regions might suffer from systematic uncertainty stemming from the uncertainty of our phenomenological model of the residual cosmic-ray background in the Pass 8 Fermi/LAT data. 
}


\maketitle

\section{Introduction}

After ten years of operations, Fermi Large Area Telescope (LAT) \citep{atwood09} has accumulated statistics of the \gr\ signal from the sky sufficient for exploration of diffuse sky emission in the TeV band, which overlaps the energy band that is accessible to the ground-based \gr\ telescopes. Although the effective collection area of the LAT is orders of magnitude smaller than that of the ground-based \gr\ telescopes, its sensitivity for the diffuse sky signal is comparable to or better than that of the ground-based telescopes because the suppression of the charged cosmic-ray background on top of which the \gr\ signal is detected is orders of magnitude better.  

Detection of diffuse \gr\ flux in the multi-TeV range has been reported by High Energy Stereoscopic System (HESS) \citep{hess}, Milagro \citep{milagro}, and the Astrophysical Radiation with Ground-based Observatory at YangBaJing (ARGO-YBJ)  \citep{argo} collaborations. Ground-based telescopes are better suited for measuring signals from isolated point sources, for which the level of cosmic-ray background can be directly estimated from a comparison of the signal from the source direction with the signal from adjacent sky regions around the source, or in the same declination strip on the sky. In contrast, measuring diffuse emission is challenging because it is impossible to find an adjacent signal-free region on the sky from a priori considerations. In this respect, even though the statistics of the Fermi/LAT signal is much lower than that of the ground-based telescopes, LAT measurements are complementary to the ground-based measurements and could be used for improvement of the ground-based measurements, for example, through identification of optimal regions for a background estimate.  

The diffuse emission in the TeV energy range comes almost exclusively from cosmic-ray interactions in the interstellar medium because the extragalactic \gr\ flux is strongly suppressed by the pair production on extragalactic background light. The inverse Compton flux from cosmic-ray electrons in the interstellar medium is suppressed by the softening of the electron spectrum in the multi-TeV range and by the onset of Klein-Nishina suppression of the Compton-scattering cross-section. Thus, the dominant component of the multi-TeV diffuse emission is provided by interactions of cosmic-ray protons and nuclei with energies in the 10-100 TeV range.  

In this respect, the study of diffuse \gr\ flux in the multi-TeV range provides a straightforward probe of the distribution of cosmic rays with energies above 10 TeV in the local interstellar medium and in the large-scale cosmic-ray halo of the Milky Way (their interactions produce diffuse emission at high Galactic latitude) and across the Milky Way disc (generating the bulk of emission at low Galactic latitude). Different models of the cosmic-ray population in the Galactic disc were considered. Lower energy Fermi/LAT data indicate that the spectrum of cosmic rays in the inner Galactic disc is harder than that in the local Galaxy (with the slope close to $dN/dE\propto E^{-\Gamma}$ with $\Gamma\simeq 2.4...2.5$, rather than $\Gamma>2.7$ measured locally) \citep{neronov_malyshev,yang,Acero_2016}. This could be explained either by a model in which the "universal" slope of the cosmic-ray spectrum  determined by the balance of injection by shock acceleration produces an injection spectrum with a slope $\Gamma_0\simeq 2,$ followed by the escape through the magnetic field, with a Kolmogorov turbulence spectrum resulting in the softening of the spectrum down to  $\Gamma\simeq \Gamma_0+\delta$, $\delta=1/3$ \citep{neronov_malyshev}, or by a model in which the energy dependence of the cosmic-ray diffusion coefficient $\delta$ changes with the distance from the Galactic center \citep{gaggero}. Model predictions of the universal and Galactocentric-distance-dependent slope  for the  diffuse emission in the TeV range are rather different, as highlighted by \citet{lipari,villante}. In the universal cosmic-ray spectrum model, the softer spectrum of the outer Galaxy and high Galactic latitude emission could in principle be explained by the influence of individual local cosmic-ray sources, which contribute sizeably to the cosmic-ray population around the Sun \citep{2Myr_15,2Myr,vela} and distort the spectrum of diffuse emission in a limited energy range and in a limited sky region.
In contrast, in the model of distance-dependent cosmic-ray spectrum, the propagation regime of cosmic rays through the interstellar medium changes because the structure of the turbulent Galactic magnetic fields changes systematically. This determines the energy-dependent diffusion of cosmic rays. 

The potential of the highest energy measurements of the sky emission by the LAT has been explored by \citet{neronov18}, who reported the first measurement of the large-scale diffuse flux of the TeV sky. This measurement used the  {\tt P8R2\_ULTRACLEANVETO\_V6} event selection of LAT events, which was characterised by the lowest  residual cosmic-ray background, when it was considered as a direct implementation to a similar Pass 7 event selection \citep{fermi_igrb}. However, no detailed information on the residual cosmic-ray background level in {\tt P8R2\_ULTRACLEANVETO\_V6} was available. Knowledge of the level of residual cosmic-ray background is important for studying diffuse TeV emission because in this energy range the cosmic-ray background starts to contribute significantly to the event sample.

The Fermi / LAT collaboration has recently released the new photon selection {\tt P8R3\_SOURCEVETO\_V2} \citep{sourceveto_bruel}, which provides the suppression of the charged cosmic-ray background comparable to that of the {\tt ULTRACLEANVETO} class, but has larger event statistics, comparable to {\tt P8R2\_CLEAN\_V6} \citep{sourceveto_bruel,fermi_1812}.  Motivated by this improvement, we performed a more advanced analysis of the hard TeV diffuse emission, which we report below. The higher signal statistics and lower level of residual cosmic-ray background enable studying the spatial morphology of the signal and better characterising its spectral properties. The better data quality enables a direct comparison with the ground-based \gr\ telescope measurements. We perform this comparison in the following sections. We also recalculate the properties of the high Galactic latitude diffuse emission, after characterising the residual cosmic-ray background in the  {\tt P8R3\_SOURCEVETO\_V2} event selection.

\section{Data analysis}

\subsection{Fermi/LAT}

Our analysis is based on Fermi/LAT data that were collected within the time interval 2008 October 27 to 2019 June 20 (we excluded the first month of Fermi/LAT operations when the veto was not operating properly\footnote{https://fermi.gsfc.nasa.gov/ssc/data/analysis/LAT\_caveats.html}). We filtered the data to retain only events belonging to the {\tt  P8R3\_SOURCEVETO\_V2} class \citep{sourceveto_bruel}, which has the best quality of the residual cosmic-ray background rejection. We processed the LAT event list using the {\it gtselect}-{\it gtmktime} chain to remove photons {\tt zmax=100} and  {\tt (DATA\_QUAL>0)\&\&(LAT\_CONFIG==1)} following the recommendations of the Fermi/LAT team\footnote{https://fermi.gsfc.nasa.gov/ssc/data/analysis/}.

Next, we divided the \gr\ event set into two parts: one for the diffuse emission components and another for the resolved sources listed in the 4FGL catalogue \citep{fermi_catalog}. To do this, we collected photons within circles of radius $0.5^\circ$ around the 4FGL sources and estimated the level of diffuse background within these circles by counting the number of photons per steradian in the parts of the sky outside the $0.5^\circ$ circles. We then calculated the cumulative 4FGL source flux within the $0.5^\circ$ circles by subtracting the estimated background from the photon counts. Finally, we estimated the total 4FGL source flux by correcting for the fraction of the signal contained within the $0.5^\circ$ radius as a function of energy. We estimated this fraction based on the radial profiles of the photon distribution around bright sources Crab, Geminga, and Mrk 421. 

To calculate the spectra of cumulative point source flux and diffuse fluxes in different parts of the sky, we calculated the exposure using the {\it gtexpcube2} routine for the exposure map in 14 logarithmically equally spaced energy bins between 1 GeV and 3.16 TeV (the highest energy to which the LAT response is calculated).  

\subsubsection{Estimating the residual cosmic-ray background}
\label{sec:cr}

The level of photon fluxes that we aim to explore is so low that the contribution of the residual cosmic-ray background into the signal could possibly not be neglected. With this in mind, we used the methods developed by \citet{fermi_igrb,sourceveto_bruel,neronov_igrb} to extract an estimate of the level of residual cosmic-ray background in the {\tt SOURCEVETO} event selection as a function of energy. 

We first estimated the contamination of the {\tt SOURCEVETO} events by the residual cosmic rays in three energy bins for which the information is implicitly given by \cite{sourceveto_bruel}: 25-40 GeV, 80-125 GeV, and 250-400 GeV. \cite{sourceveto_bruel} list the residual cosmic-ray fraction for the {\tt SOURCE} class events for the high-latitude diffuse (HLD) sky region, which corresponds to the Galactic latitude $|b|>20^\circ$, excluding circles of radius $0.2^\circ$ around 3GFL sources\footnote{We used the third version of the Fermi/LAT catalogue in the residual cosmic-ray fraction analysis to be consistent with analysis of \citet{sourceveto_bruel}, but we use the current 4FGL catalogue in the original analysis in the following sections.} and excluding the region occupied by the Fermi bubbles, which we assumed to be within the Galactic longitude $-45^\circ<l<45^\circ$. We used the same sky region in our analysis. The HLD event statistics is compared to that of reference (REF) pure gamma events collected from circles of radius $0.2^\circ$ around 3GFL sources at Galactic latitudes $|b|>20^\circ$, in the central Galaxy region at $|l|<90^\circ, |b|<5^\circ,$ and from the Earth limb in a zenith angle range $111^\circ<Zd<113^\circ$. We provide a detailed demonstration of the very low contamination of the REF event sample by residual cosmic-ray background in Appendix \ref{app}. 

\citet{sourceveto_bruel} have calculated the signal in the anti-coincidence detector of the LAT ($S_{tile}$) for REF and all HLD events belonging to the {\tt SOURCE} class. We used these results to determine the total number of all HLD events, $N_{HLD,S}$, and REF events, $N_{REF,S}$, and also their difference, $N_{CR,S}=N_{HLD,S}-N_{REF,S}$. This enables estimating the residual cosmic-ray fraction in the {\tt SOURCE} event selection in the HLD region,
\begin{equation}
    \alpha=\frac{N_{CR,S}}{N_{HLD,S}}
.\end{equation}
The values of $\alpha$ for the three energy bins are given in Table \ref{tab:source}. 

\begin{table}
    \begin{tabular}{|l|l|l|l|l|l|l|l|}
    \hline
    Energy, GeV     & $\alpha$ & $\beta$ \\
    \hline
       25-40   &$0.16\pm 0.008$ & $0.95\pm 0.02$ \\ 
       80-125   & $0.29\pm 0.02$& $0.93\pm 0.05$ \\ 
       250-400   &$0.59\pm 0.05$ & $0.90\pm 0.12$ \\
       \hline
    \end{tabular}
    \caption{Fraction of residual cosmic-ray background events in the SOURCE class in three reference energy bins, estimated based on \cite{sourceveto_bruel}. }
    \label{tab:source}
\end{table}

\begin{figure}
    \includegraphics[width=\linewidth]{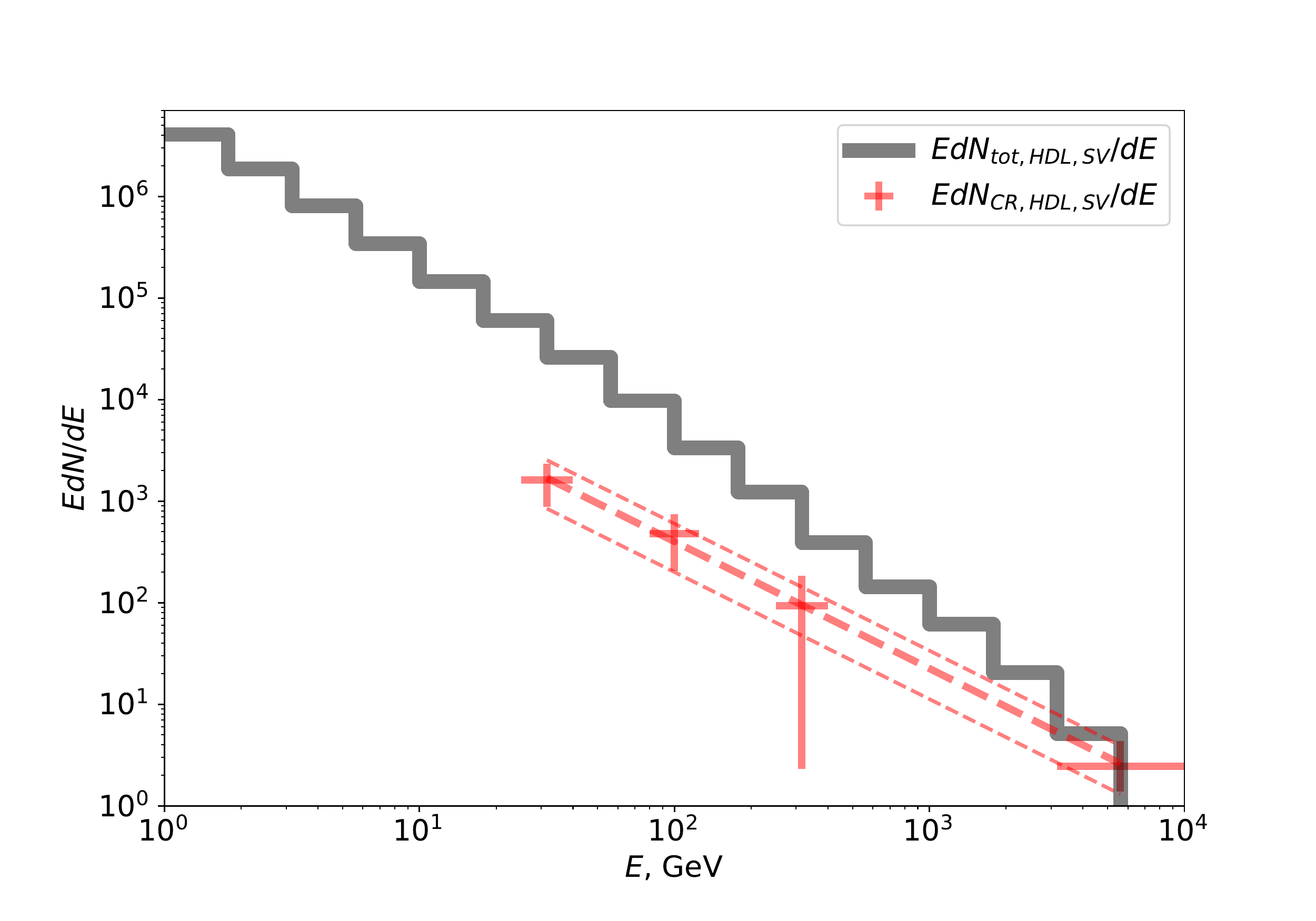}
    \caption{Residual cosmic-ray background counts $dN_{CR,HLD,SV}/dE$ in the HLD region for the {\tt SOURCEVETO} event selection (red data points and power-law fit) compared to the total counts $dN_{HLD,SV}/dE$ as a function of energy (grey histogram). 
    }
    \label{fig:cr}
\end{figure}

Knowing the number of residual cosmic-ray events $N_{CR,S}$ in the {\tt SOURCE} class event selection, we calculated the number of residual cosmic-ray events in the {\tt SOURCEVETO} selection in the same HLD region,  $N_{CR,SV}$, using the method of  \cite{neronov_igrb}. Following this approach, we first compared the statistics of events $N_{REF,S}$, $N_{REF,SV}$ of the {\tt SOURCE} and {\tt SOURCEVETO} classes in the REF samples, that is, the pure photon events in each event class. The resulting ratio
\begin{equation}
    \beta=\frac{N_{REF,SV}}{N_{REF,S}}
\end{equation}
is given in the third column of Table \ref{tab:source}. 
The measurement of $\beta$ has  allowed us to calculate the number of photon events in the HLD photon sample for the {\tt SOURCEVETO} class,
\begin{equation}
    N_{\gamma,SV}=\beta N_{\gamma,S}=\beta(1-\alpha)N_{HLD,S}
.\end{equation}
Finally, we estimated the number of residual cosmic-ray events in the {\tt SOURCEVETO} event sample in the HLD region by subtracting the photon event counts from the total event counts in the HLD region,
\begin{equation}
     N_{CR,SV}=N_{HLD,SV}-N_{\gamma,SV}
.\end{equation}
Measurements of $N_{CR,SV}$ in the three energy bins of \cite{sourceveto_bruel} are shown in Fig. \ref{fig:cr} by the red data points.

To estimate the residual cosmic-ray background at energies different from those of the three energy bins discussed by \cite{sourceveto_bruel}, we used the results of  \cite{fermi_igrb} on the residual cosmic-ray background rate spectrum, $dN_{CR}(E)/dE$, which is a power-law function of the energy in the energy range of interest (it is the rate, rather than physical flux, which is a power law in energy). This assumption is consistent with the measurement of $N_{CR,SV}$ for the three reference energies derived above, as Fig. \ref{fig:cr} shows. We added one more data point in the energy range 3-10 TeV by assuming that the totality of the counts in the HLD region at this energy belongs to the residual cosmic-ray background (there are three events in the sample). If this is not the case, the total count statistics in this energy bin provides at least an upper limit on $N_{CR,SV}$ in this energy range. 

Fitting the power-law model to the measurements at the four energies, we find the differential count rate spectrum $dN_{CR,SV}(E)/dE$. The residual cosmic-ray background contamination of the event sample could be expressed in terms of an equivalent  isotropic sky flux, if the event counts are divided by the \gr\ exposure (expressed in cm$^2$s) for the HLD sky region, even though the cosmic rays, strictly speaking, do not constitute part of the flux from the sky. In this representation, the residual cosmic-ray background level is shown in Fig. \ref{fig:spectrum_gp}. As a cross-check of the correctness of our estimate of the residual cosmic-ray background, we show in the same figure the calculation of the isotropic background template for the {\tt P8R3\_SOURCEVETO\_V2} event selection calculated by \citet{sourceveto_bruel}. The isotropic background includes both the isotropic \gr\ background (IGRB) derived by \citet{fermi_igrb} and the residual cosmic-ray background. Below an energy of 300 GeV, the isotropic background is dominated by the IGRB contribution. Above this energy, the residual cosmic-ray background dominates. The estimate of the isotropic background by \citet{sourceveto_bruel} is within the uncertainty range of our estimate of the residual cosmic-ray background in this energy range.

\section{Results}

\subsection{TeV diffuse emission from the Galactic plane}

The strongest TeV signal comes from the Galactic plane,  which contributes about $60\%$ of the all-sky signal.

\begin{figure*}
    \includegraphics[width=\linewidth]{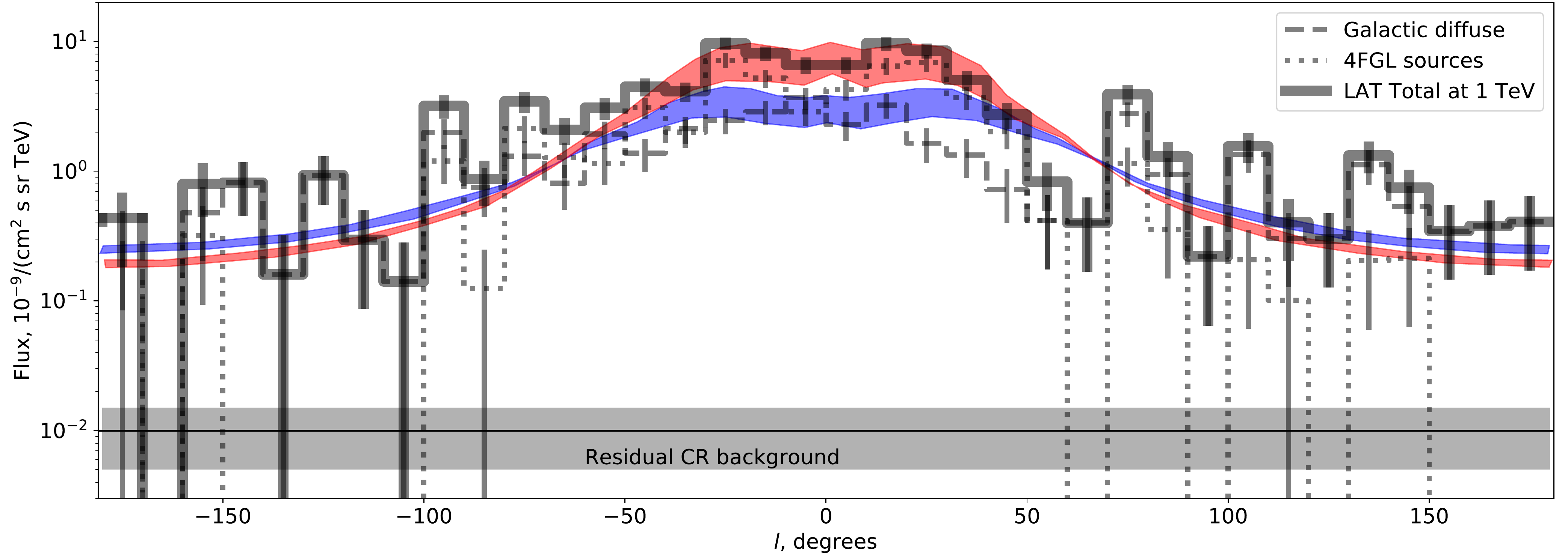}
    \caption{
    Galactic longitude profile of the signal from Galactic latitude range $|b|<2^\circ$. Dashed and dotted lines show the diffuse and resolved 4FGL source flux components. Solid line is the sum of the two components.  Blue and red model curves are from \citet{villante}. Grey band shows the level of residual cosmic ray background. }
    \label{fig:allsky_image}
\end{figure*}

The Galactic longitude profile of the signal is shown in  Fig. \ref{fig:allsky_image}. The signal is collected in bins spanning $10^\circ\times 10^\circ$ within a $|b|<2^\circ$ strip around the Galactic plane in the energy range from 0.5 TeV to 2 TeV. The signal is detected above the residual cosmic-ray background level (the equivalent flux is $10^{-11}$~/(TeV cm$^2$s sr)) at all Galactic longitudes. Emission from 4FGL sources dominates the overall Galactic plane flux within $|l|<50^\circ$ of the inner Galactic plane part, the diffuse emission provides the dominant flux component in the outer Galaxy.

\begin{figure}
    \includegraphics[width=\linewidth]{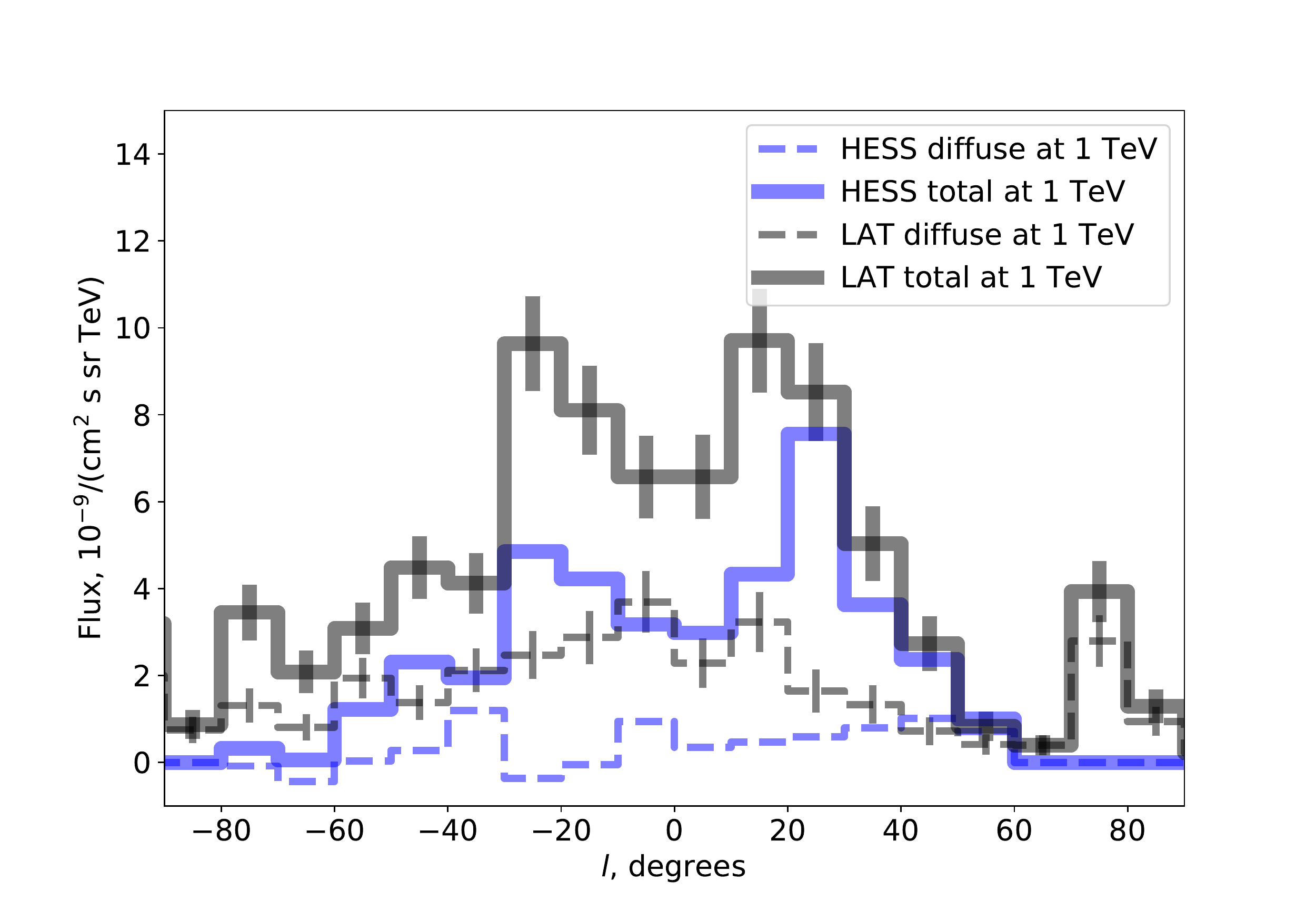}
    \includegraphics[width=\linewidth]{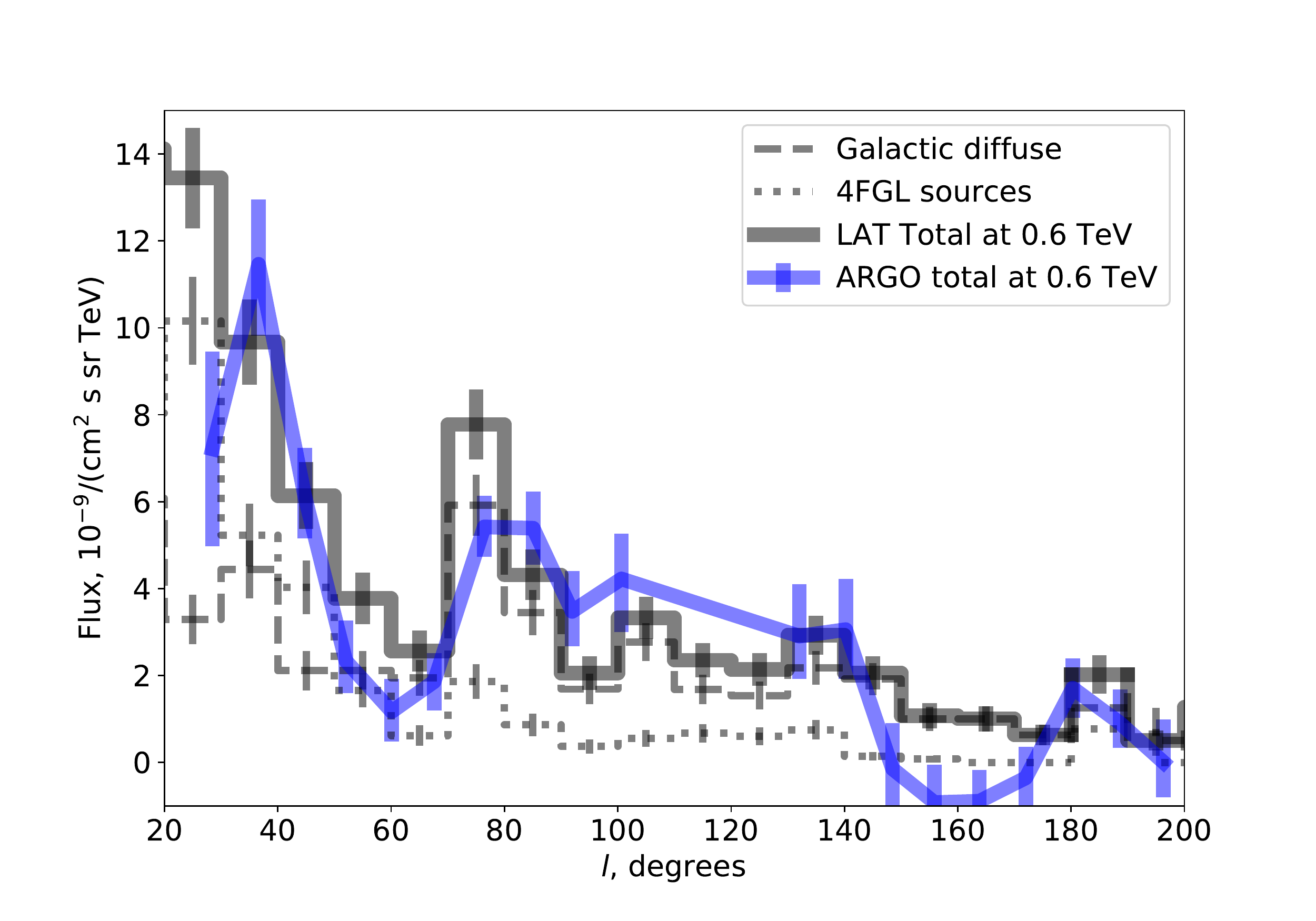}
    \caption{Top: Fermi/LAT (black) and HESS (blue)  measurements of the Galactic plane flux at 1~TeV energy within a strip $|b|<2^\circ$. Bottom: Fermi/LAT (black) and ARGO-YBJ (blue) measurements of the Galactic plane flux within the $|b|<5^\circ$ strip in the energy range around $0.6$~TeV. Notations are the same as in Fig. \ref{fig:allsky_image}.  Dashed lines show diffuse emission, and solid lines are the total flux (diffuse emission plus resolved sources). }
    \label{fig:GP_comparison}
\end{figure}

\begin{figure}
    \includegraphics[width=\linewidth]{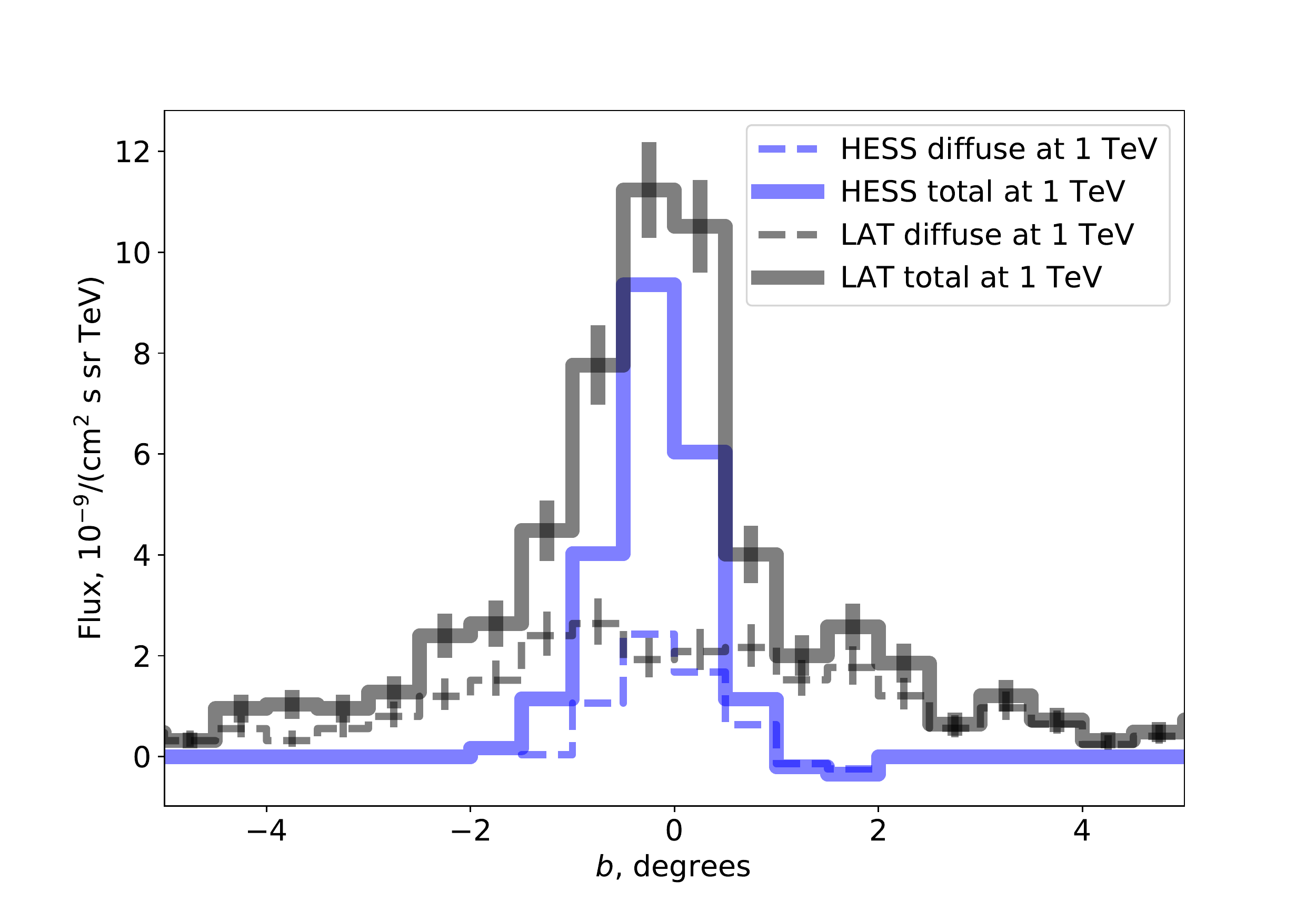}
    \caption{Fermi/LAT and HESS measurements of the Galactic latitude profile of the inner Galactic plane strip $-80^\circ<l<60^\circ$. Notations are the same as in Fig. \ref{fig:GP_comparison}.}
    \label{fig:profile_b}
\end{figure}

Detection of the diffuse Galactic plane signal at TeV has previously been reported by the HESS \citep{hess} and ARGO-YBJ \citep{argo} collaborations. Figs. \ref{fig:GP_comparison} and \ref{fig:profile_b} provide a comparison of Fermi/LAT measurements with these previous measurements. 

There is a significant difference between the measurements of Fermi/LAT and HESS (which should be considered as a lower bound on the flux \citep{hess}). The overall (resolved source $+$ diffuse emission) flux measured by HESS in the $-80^\circ<l<60^\circ$ part of the Galactic Plane within the Galactic latitude range $|b|<2^\circ$ is almost everywhere lower than the Fermi/LAT measurements of the flux from the same sky region. The only exception is the region $30^\circ<l<60^\circ$ , where the two flux measurements are compatible. 

The agreement between Fermi/LAT and ARGO-YBJ \citep{argo} measurement of the flux within the $|b|<5^\circ$ strip is better. The two measurements are consistent in large parts of the Galactic plane, except for the Cygnus region at $l\sim 80^\circ$ , where the LAT detects much higher flux and in the outer Galaxy part of the Galactic plane, $140^\circ<l<170^\circ$, where ARGO-YBJ does not detect any diffuse emission flux, while Fermi/LAT has significant flux detection. 

The separation of the total flux into diffuse and resolved source components strongly depends on (a) the number of resolved sources in each telescope and (b) the angular cut on the source extent for the extended sources. Most of the sources in the inner Galactic plane are extended, and the extensions measured by HESS are different from those measured by Fermi/LAT \citep{tev_cr_sources}. Because of this fact, a stronger discrepancy is noted between the Fermi/LAT and HESS  measurements of the diffuse emission component. The diffuse emission flux measured by the LAT has a level that is comparable to the total flux measured by HESS, while the HESS estimate of the diffuse emission is almost an order of magnitude lower. 

We attribute the discrepancies between the Fermi/LAT and HESS measurements to the peculiarity of the background subtraction in the HESS data. First, the HESS source detection method relies on the ring background estimate method in which the background level is judged based on the count statistics in ring segments around the reference point at which the \gr\ signal is estimated. This assumes that there is no \gr\ emission at the background ring position, which is not correct in the case of the signal of the Galactic plane. 

Next, the HESS analysis of the diffuse emission assumed that there is no diffuse emission signal outside the $|b|<1.5^\circ$ strip. Fig \ref{fig:profile_b} shows that this is not the case. The Galactic latitude profile of the HESS signal has to converge to zero by construction at $|b|=1.5^\circ$. The Fermi/LAT analysis does not rely on the assumption of the absence of the signal at $|b|>1.5^\circ$, and indeed, the signal is not zero in this region. The signal that is measured by Fermi/LAT at $|b|>1.5^\circ$ is counted as part of the background in the HES analysis. This leads to an over-estimate of the background level. The difference between the Fermi/LAT measurements and the HESS lower bounds is more sophisticated in the Galactic longitude profiles, but the origin of the difference remains the same. The over-estimate of the background in the HESS measurements depends on the Galactic longitude, so that the discrepancy between Fermi/LAT and HESS also depends on the Galactic longitude. This explanation of the discrepancy between the Fermi/LAT and HESS profiles is supported by the fact that    the level of diffuse flux measured by  Fermi/LAT in the HESS  background estimate regions is comparable to the overall mismatch between the Fermi/LAT and HESS diffuse flux measurements within the $\pm 2^\circ$ strip around the Galactic plane. 

The ARGO-YBJ background estimate method is different and by construction less sensitive to the details of the \gr\ signal distribution in the immediate vicinity of the Galactic lane. ARGO-YBJ estimates the background in strips of constant declination, which mostly contain regions of high Galactic latitude, where the \gr\ flux level decreases significantly.  

The TeV \gr\ signal for the Galactic plane provides information on the distribution of cosmic rays with energies $E>10$~TeV in the Galaxy \citep{neronov_malyshev,yang,Acero_2016}. Uncertainties in the cosmic-ray source distribution throughout the Galactic disc and uncertainties in the details of cosmi- ray diffusion out of the disc lead to large uncertainties in modelling the diffuse emission flux throughout the Galactic plane. This is illustrated in  Fig. \ref{fig:allsky_image}, where recent models of the TeV Galactic plane emission calculated by \citet{villante} are shown. The level of diffuse emission is approximately reproduced in the inner Galactic plane by a model that assumes that the cosmic-ray spectral slope is harder than the locally observed slope in the inner Galaxy. However, as we discussed above, the separation of the total emission into diffuse and source components strongly depends on the assumptions about the nature and morphology of the detected sources. A large part of the detected sources might be tracing the injection points of cosmic rays \citep{tev_cr_sources}, and in this respect, they should be considered as part of the emission from cosmic-ray interactions. In this case, the model with the distance-independent cosmic-ray injection spectrum traces the TeV flux better.

\begin{figure}
    \includegraphics[width=\linewidth]{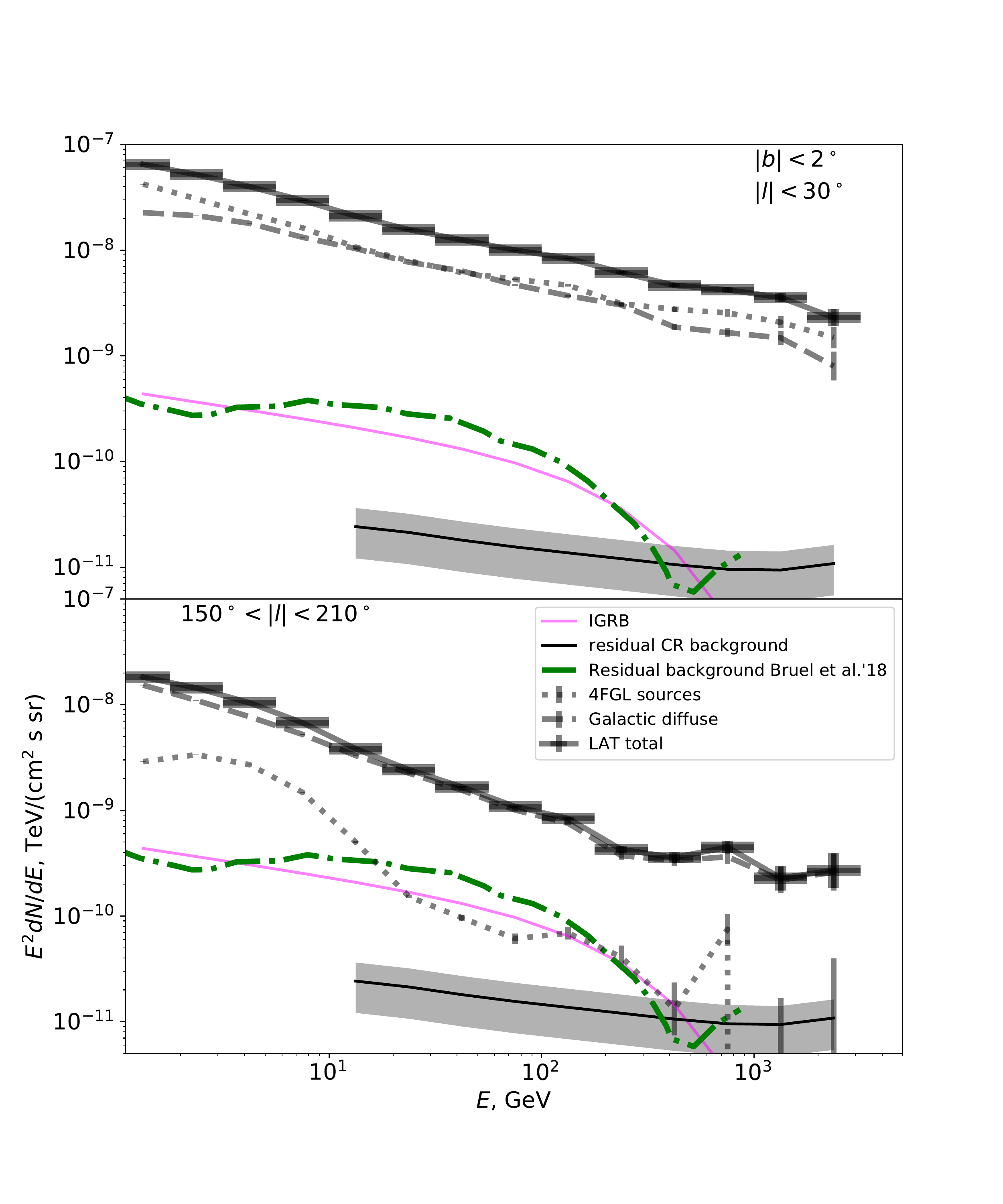}
    \caption{Emission spectrum from the low Galactic latitude region $|b|<10^\circ$. Top and bottom panels refer to the inner,  $|l|<30^\circ$, and outer, $150^\circ<l<210^\circ$,  Galaxy  segments.  Dashed and dotted lines show the diffuse background and resolved 4FGL source components of the flux. Solid lines show the sum of the two components. The thin solid magenta line shows the IGRB reported by \citet{fermi_igrb}. For comparison, the spectrum of the isotropic background template for the  {\tt P8R3\_SOURCEVETO\_V2} event selection, calculated by \citet{sourceveto_bruel}, is shown by the green dash-dotted line. This isotropic background template includes the IGRB and the residual cosmic-ray background components.}
    \label{fig:spectrum_gp}
\end{figure}

Models with a distance-dependent and a universal cosmic-ray spectrum in the Galactic disc also predict different shapes of the spectrum in different directions along the disc \citep{lipari}. Fig. \ref{fig:spectrum_gp} shows a comparison of the spectra of diffuse emission from the Galactic ridge ($|l|<30^\circ$) and the outer Galactic plane ($150^\circ<l<210^\circ$). In the energy range between 30 and 300 GeV, the two spectra clearly have different slopes. This could be well explained by the phenomenological model of the Galactic distance-dependent slope of the average cosmic-ray spectrum \citep{gaggero}, and also by the model of the universal hard cosmic-ray spectrum, with the effect of the local source imprinted on the outer Galaxy spectrum \citep{neronov_malyshev,2Myr,vela}. However, the TeV band spectra of the two regions have consistent slopes: $\Gamma=2.38\pm 0.12$ for the Galactic ridge, and $\Gamma=2.23\pm 0.16$ for the outer Galactic plane.  This is difficult to explain in the model of \cite{gaggero}, in which the outer Galactic disc spectrum has to remain soft in the TeV range as well. 

The hard spectrum of the outer Galactic disc is consistent with the model of anisotropic cosmic-ray diffusion, which reconciles the measurement of the structure of the Galactic magnetic field and cosmic-ray data \citep{Giacinti:2017dgt}. Within this model, a small number of cosmic-ray sources provides a sizeable fractional contribution to the overall local cosmic-ray population and to the local \gr\ emissivity of the interstellar medium \citep{2Myr,vela}. In contrast, due to the projection effects, many more sources contribute to the spectrum of the Galactic disc, so that its slope provides a measurement of the average slope of the Galactic cosmic-ray  population \citep{neronov_malyshev}. 

\subsection{High Galactic latitude emission}

\begin{figure}
    \includegraphics[width=\linewidth]{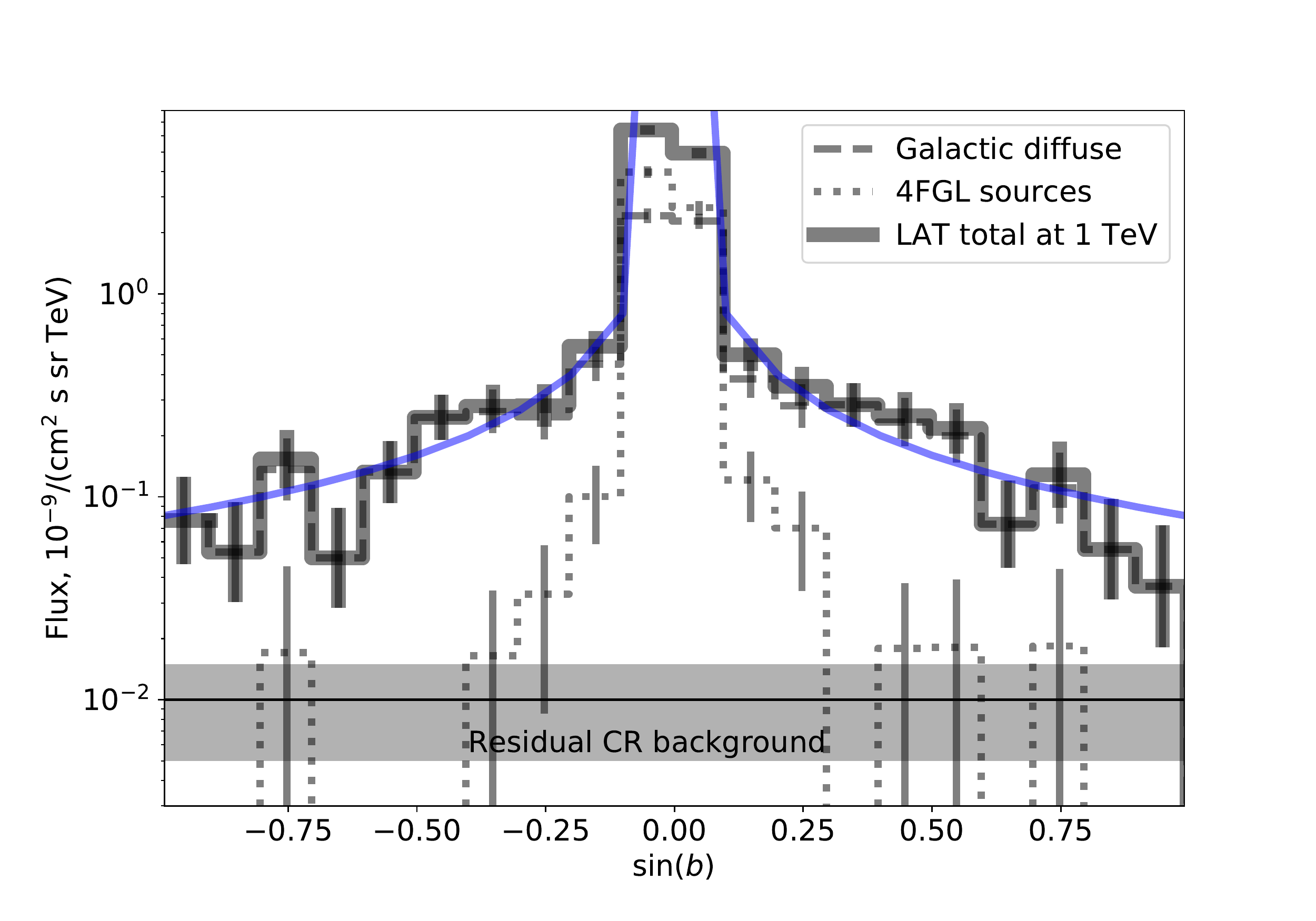}
    \includegraphics[width=\linewidth]{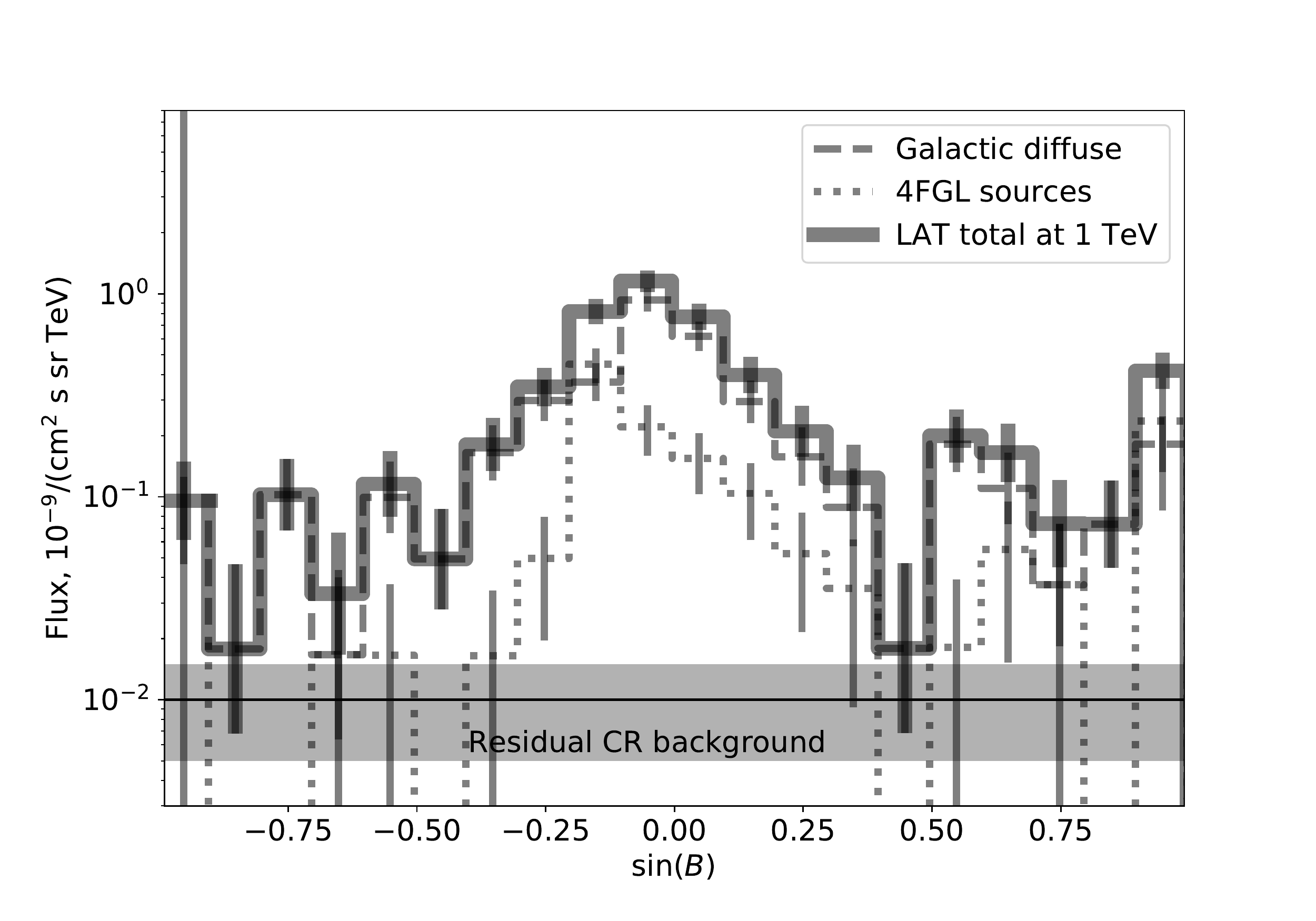}
    \caption{Galactic latitude profiles of the inner Galactic plane strips $-60^\circ<l<60^\circ$ (top panel) and  $120^\circ<l<240^\circ$ (bottom panel). The blue curve shows the emission from a disc with constant thickness. The other notations are the same as in Fig. \ref{fig:allsky_image}.}
    \label{fig:fermi_profile_b}
\end{figure}

Although the level of diffuse Galactic emission outside the Galactic plane is much lower, its flux is still detected at a high significance level. Fig. \ref{fig:fermi_profile_b} shows the Galactic latitude profile of the signal from the same strip as in Fig. \ref{fig:profile_b}, but up to the high Galactic latitude range (binned linearly in $\sin(b)$). The emission is detected in the Galactic pole regions (the highest latitude bins span $65^\circ<|b|<90^\circ$), well above the residual cosmic-ray background level. The high Galactic latitude profile is consistent with the simple model of emission from a homogeneous disc of constant thickness. In this model the signal is proportional to the column density of the disc, which scales as $1/\sin|b|$ with Galactic latitude. This model is shown by the blue line in the top panel of Fig. \ref{fig:fermi_profile_b}. 

The simple constant-thickness disc model does not fit the profile of the TeV signal in the outer Galaxy direction in the sector $120^\circ<l<240^\circ$, as the bottom panel of Fig. \ref{fig:fermi_profile_b} shows. The central part of the disc emission at low latitude is missing and the signal has high-latitude flattening. This is perhaps explained by the truncation of the Galactic disc beyond the solar radius and/or by the presence of the complex local interstellar medium, with the Local Bubble \citep{localbubble} introducing large variations in the column density of the interstellar material in different directions.
 
\begin{figure}
    \includegraphics[width=\linewidth]{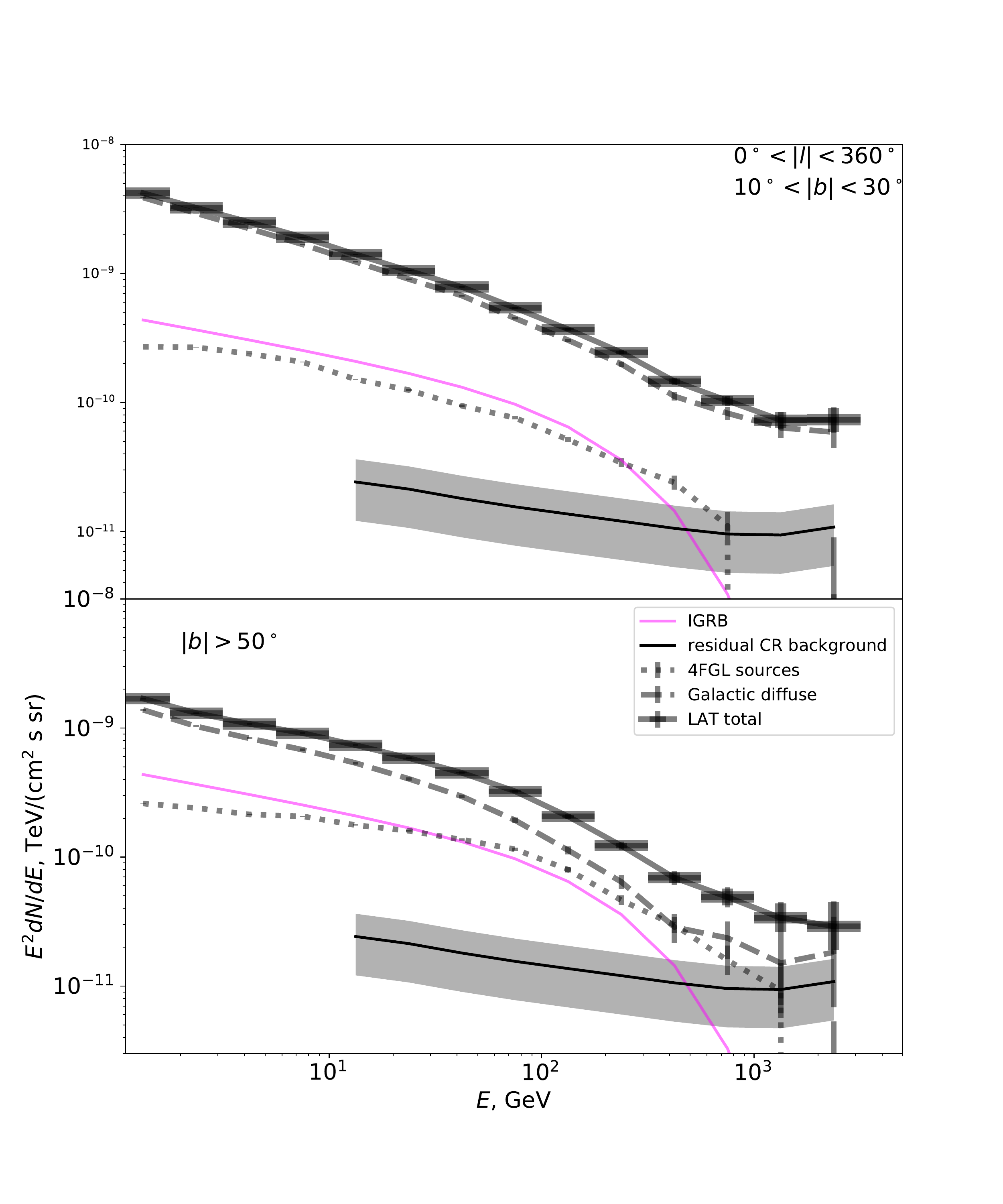}
    \caption{Spectra of high Galactic latitude regions:  mid-latitudes $10^\circ<|b|<30^\circ$ (top panel) and the Galactic pole regions $|b|>50^\circ$ (bottom panel).  The notations are the same as in Fig. \ref{fig:spectrum_gp}. }
    \label{fig:spectrum_high}
\end{figure}

The high Galactic latitude signal contains a resolved source and diffuse components. The spectra of the two components in the TeV range are significantly different, as shown in Fig. \ref{fig:spectrum_high}. The flux of the resolved source component is soft and sub-dominant, compared to the diffuse emission. The resolved source flux (dominated by distant active galactic nuclei) does not depend on Galactic latitude. Its spectral shape approximately repeats that of the IGRB \citep{neronov_igrb,fermi_igrb}. 

In contrast, the diffuse emission level depends on the Galactic latitude. It is interesting to note that the slopes of the 0.3-3 TeV diffuse emission spectrum at different Galactic latitudes are consistent with the slope of the Galactic plane. In the latitude range $|b|>50^\circ$ , the slope measurement is $\Gamma=2.39\pm 0.13,$ and in the $10^\circ<|b|<30^\circ$ region, it is $\Gamma=2.43\pm 0.06$.  Combining the measurements of the 0.3-3 TeV spectral slopes in four different parts of the sky (inner and outer Galactic plane, the Galactic poles, and the mid-latitude regions), we find an average slope $\Gamma=2.40\pm 0.05$ with which the spectra of all the four regions are consistent.  

\section{Discussion}

The new Fermi/LAT event selection {\tt P8R3\_SOURCEVETO\_V2} is characterised by a very low level of residual cosmic-ray background and by relatively high statistics of the \gr\ signal in the TeV energy range. These properties of the data have allowed us to study the properties of TeV Galactic diffuse emission from different parts of the sky. 

Surprisingly, the overall level of the TeV \gr\ flux from the inner Galactic plane detected by Fermi/LAT is approximately twice higher than the flux detected by the HESS telescope from the analysis of the Galactic plane survey region \citep{hess}. We attribute this discrepancy to the subtleties of the background estimate in the ground-based Cherenkov telescopes.  Cherenkov telescopes have narrow fields of view, which complicates the task of mapping the diffuse emission on large angular scales. 

The comparison of Fermi/LAT and HESS measurements provides an indication of how the ground-based measurements can be improved. The measurement quality of diffuse Galactic plane emission with ground-based instruments will be significantly improved with the start of operation of small-size telescopes of the Cherenkov Telescope Array (CTA) \citep{sst}, which will have a much wider field of view than HESS. This will enable a more reliable estimate of the background from higher Galactic latitude regions using the ring background technique. Still, thee  \gr\ flux measurement will be significantly contaminated by the cosmic-ray and incorrectly attributed \gr\ background flux  even if the Galactic latitude range of the background regions is extended up to $|b|\sim 5^\circ$. Perhaps the best background modelling technique for the CTA is to rely on Fermi/LAT measurements in the energy range of interest and use an imaging template found from Fermi/LAT data in the analysis  of individual point and extended sources. 

Our analysis shows that diffuse emission spectrum is hard in different parts of the sky, as first noted by \citet{neronov18}, based on the analysis of {\tt ULTRACLEANVETO} Pass 7 event selection with custom cross-calibration of Fermi/LAT with ground-based telescope measurements. We have used higher statistics of the TeV band signal and better calibrations available in the {\tt P8R3\_SOURCEVETO\_V2}  event selection for a more detailed investigation of the spectral and imaging properties of the hard emission. Additionally, we were able to quantify the residual cosmic-ray background contamination of the {\tt P8R3\_SOURCEVETO\_V2} signal and to verify that the hard spectral component is not generated by this contamination. 

The slope $\Gamma=2.40\pm 0.05$ of the TeV diffuse emission in different parts of the sky is harder than that of the locally measured cosmic-ray spectrum, which changes from $\Gamma>2.8$ in the energy range below 200 GeV \citep{ams-02,pamela} to $\Gamma\sim 2.6$ in the multi-TeV range \citep{cream,nucleon} (see \citet{semikoz_review} for a recent review). It is, however,  consistent with slope of the average Galactic cosmic-ray spectrum in the inner Galaxy measured based on a study of the lower energy \gr\ diffuse emission \citep{neronov_malyshev,yang,Acero_2016}. It is surprising that this hard slope is also found in high Galactic latitude and outer Galactic disc regions. In addition to the possibility that the TeV diffuse emission spectral slope corresponds to the slope of the average Galactic cosmic-ray population, possible models of the hard TeV component include diffuse emission from the interstellar medium of the local Galaxy produced by cosmic rays that spread from a nearby source into a (local) super-bubble  \citep{superbubble,vela}, or emission from the large (100 kpc scale) cosmic-ray halo around the Milky Way \citep{aharonian_gabici} or decays of super-heavy dark matter particles \citep{dm_decay,dm_decay1,Kachelriess:2018rty}.

It was noted by \citet{neronov14,neronov16,evidence,neronov18} that the flux level and spectral properties of the hard diffuse multi-TeV high Galactic latitude emission detected by Fermi/LAT are compatible with those of the IceCube neutrino signal from different parts of the sky either in the high-energy starting or muon neutrino channel at much higher energies $E>100$~TeV. In this sense, the hard spectrum 0.3-3 TeV Galactic diffuse emission could be the \gr\ counterpart of the high Galactic latitude neutrino flux. Before this nature of the multi-TeV \gr\ signal can be firmly established, it is important to extend the measurements into the energy band that reaches the IceCube energy range (10 TeV). This is possible, in principle, because the LAT detects photons with energies up to 10 TeV. The Galactic plane signal is clearly identifiable in the sky map between 3 and 10 TeV. However, the instrument characteristics of the LAT are not known because of the absence of Monte Carlo modelling of the instrument response similar to that reported by \citet{sourceveto_bruel}. This modelling has also to include the modelling of the residual cosmic-ray background that contaminates the signal more strongly in this energy range. Further improvement of the statistics of the space-based measurements of diffuse emission in the multi-TeV band should be possible with a larger space-based \gr\ telescope, such as High Energy Radiation Detector (HERD) \citep{herd}. 

A complementary probe of the hard component of Galactic \gr\ flux in the multi-TeV band is also possible with a dedicated ground-based \gr\ detector providing sufficiently strong suppression of the charged cosmic-ray background. This could be achieved by measuring the muon content of extensive air showers, through observations with Cherenkov telescopes at large zenith angle \citep{highz}, or using underground muon detectors, as demonstrated by the KArlsruhe Shower Core and Array DEtector  (KASCADE) experiment \citep{KASCADE} and as planned in the Carpet-3 detector \citep{carpet}. 

\appendix
\section{Residual cosmic-ray background contamination of the REF event sample}
\label{app}

The REF  event sample introduced by \cite{sourceveto_bruel}  is almost exclusively composed of \gr\ events in the three energy bins considered in the analysis reported by \cite{sourceveto_bruel} . An upper limit on the residual cosmic-ray contamination of this sample could be readily derived from a comparison of the statistics of REF events with that of events from the  sky region that might have the highest contamination by the residual cosmic rays: the Galactic poles at $|b|>80^\circ$. 

This comparison is given in Table \ref{tab:ref_sample}. In this table, the statistics of different contributions to the REF sample is summarised. $N_{source}$ denotes the events within the $0.2^\circ$ circles around the 3FGL sources at Galactic latitudes $|b|>20^\circ$, $N_{Gal}$ is the number of events from the inner Galactic disc $|l|<90^\circ$, $|b|<5^\circ$, and $N_{limb}$ is the number of events from the direction of the Earth limb at zenith angles $111^\circ<Zd<113^\circ$.

We derive an upper limit on the number of residual cosmic-ray events in the REF sample, $N_{CR}$,  from the event statistics in the north and south Galactic poles at $|b|>80^\circ$, which is shown in Table \ref{tab:poles_sample}. The total event counts in these regions, $N_{tot}$, have four contributions: the resolved point source counts $N_{source}$, the isotropic \gr\ background $N_{igrb}$, the  Galactic diffuse emission $N_{diff}$ , and the residual cosmic-ray background $N_{CR,poles}$.  We calculated $N_{source}$ by collecting photons from the circles of $0.2^\circ$ around known catalogue sources and correcting for the energy-dependent  fraction of the point source flux contained within the $0.2^\circ$ circles. Subtracting $N_{source}$ from the total counts, we find 
\begin{equation}
N_{CR,poles}<N_{CR,poles}+N_{IGRB}+N_{diff}=N_{tot}-N_{source}
\end{equation}
In this way we obtain a robust upper limit on $N_{CR,poles}$, rather than measurement, but this is sufficient for the demonstration of low contamination of the REF event sample by the residual cosmic rays.

After determining an upper limit on $N_{CR,poles}$ , we derived from it an upper limit on the number of residual cosmic-ray events in the REF event selection by rescaling the $N_{CR,poles}$ upper limit considering the difference in solid angle from which the signals from the Galactic poles and REF regions are collected.  The regions at  $|b|>80^\circ$ span the solid angle $\Omega_{|b|>80^\circ}\simeq 0.19$ sr. The regions of which the REF sample is composed span overall a solid angle $\Omega_{REF}\simeq 0.82$~sr.  Thus, the number of cosmic-ray events in the REF sample is 
\begin{equation}
    N_{CR}=\frac{\Omega_{REF}}{\Omega_{|b|>80^\circ}}N_{CR,poles}<\frac{\Omega_{REF}}{\Omega_{|b|>80^\circ}}(N_{tot}-N_{source})
.\end{equation}
This upper bound is given in  Table \ref{tab:ref_sample}.

\begin{table*}
    \begin{tabular}{|l|l|l|l|l|l|l|l|}
    \hline
    Energy, GeV     & $N_{source}$&$N_{Gal}$& $N_{limb}$& $N_{REF}$& $\sqrt{N_{REF}}$&$N_{CR}$&$\sqrt{N_{REF}+N_{CR}^2}/N_{REF}$ \\
    \hline
       25-40   &5524&47406&227668&280598&530&$<2071$& $0.8\%$  \\ 
       80-125   &958&8641&31626&41225&203&$<397$& $1.1\%$ \\ 
       250-400   &136&1744&4968&6848&83&$<95$& $1.8\%$  \\
       \hline
    \end{tabular}
    \caption{Statistics of events in the REF event sample of \cite{sourceveto_bruel}. }
    \label{tab:ref_sample}
\end{table*}

\begin{table}
    \begin{tabular}{|l|l|l|l|l|l|l|l|}
    \hline
    Energy, GeV     & $N_{tot}$&$N_{3GFL}$& $N_{diff}+N_{IGRB}+N_{CR,poles}$ \\
    \hline
       25-40   &784&304&480  \\ 
       80-125   &154&62&92 \\ 
       250-400   &33&11&22  \\
       \hline
    \end{tabular}
    \caption{Statistics of events in the north and south Galactic poles }
    \label{tab:poles_sample}
\end{table}

We include this upper bound as the systematic error on the measurement of the number of \gr s in the REF regions and add it in quadrature to the statistical uncertainty of   $N_{REF}$, as specified in the last column of Table \ref{tab:ref_sample}.  This uncertainty is then included in the calculation of the uncertainty of $\alpha$ reported in Table \ref{tab:source}.  

\bibliographystyle{aa}
\bibliography{sourceveto.bib}
\end{document}